\documentclass[preprint,eqsecnum,showpacs,aps,epsfig,floats,groupedaddress,superscriptaddress,showkey]{revtex4}

\usepackage{graphicx}  
\usepackage{subeqnar}  

\begin{document}

\preprint{cond-mat/0402247}

\title{Superconducting order parameter in heavily overdoped
Bi$_2$Sr$_2$CaCu$_2$O$_{8+\delta}$: a global quantitative
analysis}

\author{R. H. He\footnote[3]{Present address: Department of Applied Physics,
Stanford University, Stanford, California 94305}}
\affiliation{Physics Department, Applied Surface Physics State Key
Laboratory, \\and Synchrotron Radiation Research Center, Fudan
University, Shanghai 200433, China}
\author{D. L. Feng}
\email{dlfeng@fudan.edu.cn} \affiliation{Physics Department,
Applied Surface Physics State Key Laboratory, \\and Synchrotron
Radiation Research Center, Fudan University, Shanghai 200433,
China}
\author{H. Eisaki}
\affiliation{Nano-electronic Research Institute, AIST, Tsukuba
305-8568, Japan}
\author{J.-i. Shimoyama}
\affiliation{Department of Applied Chemistry, University of Tokyo,
Tokyo, 113-8656, Japan}
\author{K. Kishio}
\affiliation{Department of Applied Chemistry, University of Tokyo,
Tokyo, 113-8656, Japan}
\author{G. D. Gu}
\affiliation{Physics Department, Brookhaven National Laboratory,
P. O. Box 5000, Upton, New York 11973, USA}

\date{\today}

\begin{abstract}
A systematic analysis of the complex lineshape at $(\pi,0)$ of
heavily overdoped Bi$_2$Sr$_2$CaCu$_2$O$_{8+\delta}$ is presented.
We show that a coherent component in the quasi-particle excitation
is correlated with the superfluid density throughout the entire
doping range and suggest it is a direct measure of the order
parameter of high temperature superconductors.
\end{abstract}

\pacs{71.10.Ay, 74.72.Hs, 79.60.Bm}

\maketitle

The off-diagonal nature of the order parameter of a superconductor
makes it difficult to be measured directly\cite{Yang}. For
conventional superconductors, the appearance of an energy gap in
the single particle excitation spectrum characterizes the
superconducting state; as a result, the observation of the gap is
virtually equivalent to the observation of the superconducting
order parameter. On the other hand, for high temperature
superconductors (HTSC's), because of the pseudogap phenomenon in
the normal state of underdoped systems, such a strict link between
the gap and off-diagonal long range order (ODLRO) breaks. There
have been suggestions between ODLRO and the $(\pi,\pi)$ resonance
peak in the dynamic structure factor measured by inelastic neutron
scattering experiments around 40meV in HTSC's \cite{Sudip00}.
However, because a significant amount of spin fluctuations exist
in this region and evolve with temperature in the normal state of
underdoped systems, a direct and clean measurement of the
superconducting order parameter is still not achieved.

We pointed out earlier that in the single particle excitation
spectrum a sharp peak, which develops near $(\pi,0)$ of the
Brillouin zone below the superconducting transition temperature
($T_c$), exhibits similar behaviors as the superfluid density
\cite{FengScience}. This so-called superconducting peak (SCP) is
observed by angle-resolved photoemission spectroscopy (ARPES) in
many HTSC's, first on $Bi_2Sr_2CaCu_2O_{8+\delta}$ (Bi2212), later
on $YBa_2Cu_3O_{7-\delta}$, and
$Bi_2Sr_2Ca_2Cu_3O_{10+\delta}$\cite{ubiquitousPDH}. Being a probe
for single particle excitation, ARPES cannot measure collective
properties such as the superfluid density (or phase stiffness)
directly. However, it could be measuring condensate fraction, a
single particle quantity closely related to superfluid density.
Condensate fraction is defined as fraction of particles
participating in pairing, and thus is the direct measurement of
the ODLRO. Unlike the gap, the correlation between the SCP and
superconducting order parameter works well in the underdoped
regime. However, its validity in the overdoped regime has been
questioned. Some claim that the SCP increases upon overdoping,
while the superfluid density decreases\cite{DingPRL01}. Others
claim that there is no SCP on the overdoped sample, and the
peak-dip-hump (PDH) lineshape associated with the SCP is simply
due to the bilayer band splitting (BBS)
effect\cite{Barry,Fink2,Fink3,Fink4}.

In this paper, we present a global study on the photon energy
($h\nu$) and temperature ($T$) dependent ARPES spectra of heavily
overdoped Bi2212 at $(\pi,0)$. We show that there is a substantial
SCP in the heavily overdoped regime and its correlation with the
ODLRO holds across the entire doping range. This also gives a
unified understanding of the PDH lineshape in the BBS
circumstance\cite{FengPRL01}: the anti-bonding band (AB) and
bonding band (BB) each develop their own PDH lineshapes by
transferring parts of their spectral weight into the newly-created
coherent quasiparticles (AB peak and BB peak, respectively).

ARPES experiments were performed at beamline V-4 of Stanford
Synchrotron Radiation Laboratory (SSRL). Data were collected with
He-I light from a He discharge lamp (for the $T$-dependence) and
polarized synchrotron light from a normal incidence monochromator
(for the $h\nu$-dependence), where the contamination to the
spectra from the second order light is extremely weak, ideal for a
quantitative comparison between spectra taken with different
$h\nu$'s. Detailed descriptions of experimental settings and
sample preparation can be found in Ref.\cite{FengPRL01}.

The existence of the coherent peak has been discussed extensively
for optimally doped and underdoped systems based on doping and $T$
dependence\cite{FengScience}. Here we recap this with the data
taken on an optimally doped Bi2212 ($T_c=$90K) sample in Fig. 1.
As shown in Fig. 1a, the sharp coherent peak grows out of a smooth
spectrum with decreasing temperature at $(\pi,0)$. Fig. 1b shows
how this peak behaves along $\Gamma-M$. We found that the SCP is
strongest when the hump is within 120$\sim$150 meV below $E_F$;
away from this region, the SCP rapidly dies out while dispersing
slowly with the hump to higher binding energies. The extracted
superconducting state dispersions of the peak and hump in the main
band are plotted in Fig. 1c. We emphasize that the SCP is not the
AB, otherwise, one would obtain a bilayer splitting increasing
towards $(0,0)$ (Fig. 1d), while band theory as well as symmetry
arguments gives zero splitting along
$(0,0)-(\pi,\pi)$\cite{NormanEq}.

The complication in the heavily overdoped regime originates from
the resolution of the AB and BB in the normal state, which forms a
PDH-like lineshape with the sharp AB in the lowest binding energy
position\cite{FengPRL01}. Kordyuk {\sl et al.} have implemented a
model fit of two bilayer-split bands to study the $h\nu$-dependent
superconducting state $(\pi,0)$ spectra in this doping
regime\cite{Fink2}. A qualitative agreement between the fitted
$h\nu$-dependent matrix elements of the peak and hump and those of
bilayer-split bands predicted by
theory\cite{FengPRB02,LindroosPRB} was argued to justify the peak
and hump being due to the AB and BB, at least in the overdoped
regime\cite{Fink4}. However, we found that although the fitting
function is able to fit each single spectrum well at different
$h\nu$'s, it faces difficulty when more physical constraints are
included. One example is shown in Fig. 2a. These spectra are taken
at $(\pi,0)$ on an overdoped $T_c=$65K sample (OD65) at different
temperatures. We start with the nine normal state EDC's to obtain
the global parameters through a global fit (described below). The
fit gives an excellent agreement with the experiments which
confirms the qualification of the chosen form of spectral function
for describing the normal state features and, notably, the thermal
sharpening-up process on the lineshapes upon $T$ decrease.
However, with these parameters, the discrepancy between the eight
simulated and experimental superconducting state spectra increases
upon going further into the superconducting state. This
demonstrates a drastic lineshape variation beyond the description
where only two bilayer-split bands are considered. The much more
rapid intensity accumulation on the experimental peak feature
below $T_c$ requires further investigation.

Reflecting the behavior of the SCP in underdoped, optimally doped,
and even slightly overdoped samples\cite{CampuzanoPRL99}, we
extend the fitting function in Ref.\cite{Fink2} by adding two
spectral components to account for the rapid spectral weight
accumulation near $E_F$.
\begin{eqnarray}
\label{eq:Isum}
I(\omega,T,h\nu)&=&I_0(T,h\nu)\cdot[(\sum_{\alpha}^4J_{\alpha}(\omega,T,h\nu)\cdot f(\omega,T)) \nonumber \\
&&\otimes R(\omega,\Gamma^\prime(h\nu))+B(\omega,T)]+I_1(T,h\nu)
\nonumber
\end{eqnarray}
, where $f$ is the Fermi function, $R$ the $h\nu$-dependent
resolution Gaussian, $B$ the momentum-independent but
$T$-dependent empirical background function obtained by a separate
fit on the featureless energy distribution curves (EDC's) around
$(\pi/2,\pi/2)$\cite{BackgroundNote}, $I_0$ and $I_1$ the linear
intensity coefficients. The summation is over the spectral
intensity of AB hump, BB hump, AB peak and BB peak, given by,
respectively,
\begin{eqnarray}
\label{eq:I4}
J_{ah}(\omega,T,h\nu)&=&M_{ah}(h\nu)\cdot C_a(T)\cdot A_h(\omega,T,\alpha,\varepsilon_{ah}), \nonumber \\
J_{bh}(\omega,T,h\nu)&=&M_{bh}(h\nu)\cdot C_b(T)\cdot A_h(\omega,T,\alpha,\varepsilon_{bh}), \nonumber \\
J_{ap}(\omega,T,h\nu)&=&M_{ap}(h\nu)\cdot (1-C_a(T))\cdot A_p(\omega,\Gamma_a(T),\varepsilon_{ap}), \nonumber \\
J_{bp}(\omega,T,h\nu)&=&M_{bp}(h\nu)\cdot (1-C_b(T))\cdot
A_p(\omega,\Gamma_b(T),\varepsilon_{bp}) \nonumber
\end{eqnarray}
, where $M$ is the matrix element, $C$ the remaining spectral
weight in the hump, $A_h$ and $A_p$ the self-normalized spectral
functions for the hump and the peak, respectively, given by
\begin{eqnarray}
\label{eq:A}
A_h(\omega,T,\alpha,\omega_0)&=&\frac{\xi|\sum^{\prime\prime}(\omega,T)|}
    {(\omega-\sqrt{\omega_0^2+\Delta_{sc}(T)^2})^2+\sum^{\prime\prime}(\omega,T)^2},\nonumber \\
A_p(\omega,\Gamma(T),\omega_0)&=&\frac{2\sqrt{\ln2}}{\sqrt{\pi}\Gamma(T)}
    \exp[-(\frac{\omega-\omega_0}{\Gamma(T)/2})^2].\nonumber
\end{eqnarray}
We adapt the empirical
$\sum^{\prime\prime}(\omega,T)=\sqrt{(\alpha\omega)^2+(\beta
T)^2}$ from Ref.\cite{Fink2}, where $\alpha$ and $\beta$ are made
globally fittable ($h\nu$- and $T$-independent) to reflect the
doping and impurity level of samples, and $\xi$ is a normalization
factor. $\Gamma(T)$ is a $T$-dependent linewidth (FWHM). Note that
three major extensions in the fitting function have been made.
First, two spectral components for the SCP's are added with their
Guassian lineshapes justified by the STM results\cite{SCPbySTM};
second, spectral weight transfer (SWT) ($=1-C_{a(b)}, 0\leq
C_{a(b)}(T)\leq 1$) is proposed for the formation of SCP, which is
a pronounced character of many correlated systems and is also
based on the empirical observation of the fact that the integrated
spectral weight is roughly $T$-independent in the $(\pi,0)$ region
\cite{RanderiaPRL95}; third, a realistic SCG opening extracted
with the leading edge gap method on the same sample introduces an
effective $T$-dependent energy shift to the bilayer-split bands in
the superconducting state.

In order to retrieve reliable quantitative information from a
multi-dimensional fit to a set of interconnected EDC's, we
developed a non-linear least-squares fitting routine, which has
been applied in the normal state fit in Fig.2a. The goal is to
achieve a systematic understanding of both the $h\nu$- and the
$T$-dependent behavior of the lineshape. Physical constraints are
strictly followed here, for example, in the the $h\nu$-dependent
set, all functions of $T$ are global variants to all EDC's at the
same $T$ while the functions of $h\nu$ (matrix elements) are
locally specified, and in the $T$-dependent set, vice versa. The
renormalized band energy position $\varepsilon$ is always a global
fitting parameter for the EDC's of samples at the same doping
level. Without any manual tuning in the global parameters, the new
fitting function can yield a robust, physically-constrained global
fit for a large set of EDC's regardless of the initial values
input.

It has been shown before that the intensity of the BB peak drops
quickly when the BB disperses deep below the $E_F$ at $(\pi,0)$
\cite{FengPRL01}. For simplicity, we first neglect $J_{bp}$ term,
and achieve an excellent global fit for all the spectra shown in
Fig. 2b-d simultaneously, which include the aforementioned
T-dependence data and the $h\nu$-dependence $(\pi,0)$ data of a
Pb-doped OD65 (Pb-OD65) taken above and below $T_c$. In Fig. 3a,
the $T$-dependent energy positions of the maxima of three spectral
features are shown. In the superconducting state, the AB peak is
located quite close to the AB hump, which makes it very difficult
to resolve them separately in experiments. For the AB (BB) hump
maximum, the slight decrease (increase) above $T_c$ is fully
determined by the $T$-dependence of the self-energy with the
fitted $\alpha=0.715\pm0.006eV^{-1}$ and
$\beta=2.57\pm0.03eV^{-1}$ while the superconducting state is
mainly affected by the opening of the gap. The spectral maximum
shift in the BB hump is $T$-insensitive as compared with the AB
hump; for the AB peak which turns up in the superconducting state,
its almost-constant location ($-18.01\pm0.03 meV$) seems to define
the maximum SCG opening at $(\pi,0)$. This contradicts the
scenario where the peak near $E_F$ is just the AB which should
change its position with the SCG opening. The extracted
$T$-averaged BBS amplitude is $84.39\pm0.07$
meV\cite{BBSamplitude}, consistent with Ref.\cite{FengPRL01}.

The fitted matrix elements as a function of $h\nu$ for the three
spectral features are shown in Fig. 3b. Matrix elements of AB and
BB hump vary in the same range of magnitude but with an
anti-phase-like $h\nu$-dependence as predicted by
theory\cite{FengPRB02,LindroosPRB}. Note that the
self-normalization of the spectral function is required not only
for a well-defined SWT analysis but also for a physically
comparable matrix element between different spectral components.
 $<M_{bh}>/<M_{ah}>=0.966\pm0.007$, while the matrix
elements of AB peak shows a different $h\nu$-dependence with those
of the humps and ($<M_{ap}>/<M_{ah}>=0.610\pm0.007$). This
deviation is not unexpected because SCP is of redistributed
spectral weight related to the superfluid, whose wavefunction may
have different properties with the hump counterpart.

The $T$-dependent SWT from the above fit is summarized in Fig. 4a.
The most prominent feature is the superconducting peak emerges
only below $T_c$, while the SCP at lower doping could emerge 10K
above $T_c$ (see Fig. 1a and Ref.\cite{FengScience}). This may
associate with the superconducting fluctuations in the lower
doping regime\cite{XuNature,Orenstein}. The SWT of AB is
extrapolated to be $0.223\pm0.028$ at 0 K, a value far below 1,
which casts doubt on the scenario where the entire lower-lying AB
peak possesses almost all the weight of AB
quasiparticle\cite{NormanPRL02}.  We also present the result of a
4-component fit with both $J_{ap}$ and $J_{bp}$ terms. The BB peak
weight extrapolates at 0 K to be only $0.015\pm0.008$, much
smaller than that of AB, which also justifies the 3-component fit.

In Fig. 4a, the superfluid density measured by muon spin
relaxation ($\mu$SR) at the same doping level is also displayed,
which resembles the SWT in the T-dependence. In
Ref.\cite{FengScience}, we proposed a phenomenological
substraction of a smooth background to retrieve the SCP, which was
further quantified by defining the so-called superconducting peak
ratio (SPR) as the SCP intensity divided by the overall spectral
weight in $(\pi,0)$ spectrum taken at $h\nu$=22.7 eV and low
temperature. Although the SWT is more physically well-defined than
SPR in that it is free of the matrix element effects and takes
into account the multi-component lineshape, when the hump is broad
near or below optimal doping, SPR obtained by the phenomenological
procedure reasonably characterizes the SCP. Thus, we calculated
SPR for OD65 at $h\nu$=22.7 eV at $10 K$ based on the globally
fitted AB peak. It is plotted in Fig. 4b together with SPR's at
lower doping levels reproduced from Ref.\cite{FengScience}. One
can clearly see that SPR decreases with doping in the heavily
overdoped regime, similar to the superfluid
density\cite{UemuraPlot,FengScience}. To account for this
decrease, it has been suggested that the overdoped system is
phase-separated into superconducting regions and hole rich normal
metal regions\cite{UemuraPlot}.  In any case, these two intimate
correlations between the SCP and superfluid suggest the SWT is a
direct measurement of the superconducting order parameter.

We note that due to the lack of quantitative theories, the fitting
formula used here is somewhat empirical, and certainly could be
replaced by more sophisticated formulae with different number of
components. However, the essence of SWT or any other equivalent is
to capture the anomalous rapid spectral weight growth below $T_c$
at the gap energy scale beyond the Fermi-liquid-based descriptions
and to offer a direct measure of the superconducting order
parameter which, in return, facilitates examination of various
theoretical models. There have been a few theoretical proposals of
such a quasiparticle peak created in the superconducting state by
spectral weight transfer\cite{Theories}, and its weight was
related to the condensate fraction or ODLRO. However, most of
these theories are based on underdoped cuprates, while the
overdoped regime is seldom addressed. The observed correlation
between the SWT and superfluid density throughout the entire phase
diagram calls for further theoretical efforts.

We thank Dr. D. H. Lu, and W. S. Lee at SSRL for experimental
help, and  K. M. Shen  and Dr. A. Damacelli for comments and
discussions. G.D.G. is supported by Department of Energy under
contract No. DE-AC02-98CH10886. D.L.F. is supported by the
National Science Foundation of China. SSRL is operated by the DOE
Office of Basic Energy Science Divisions of Chemical Sciences and
Material Sciences.



\begin{figure}[t!]
\centerline{\includegraphics[width=3in]{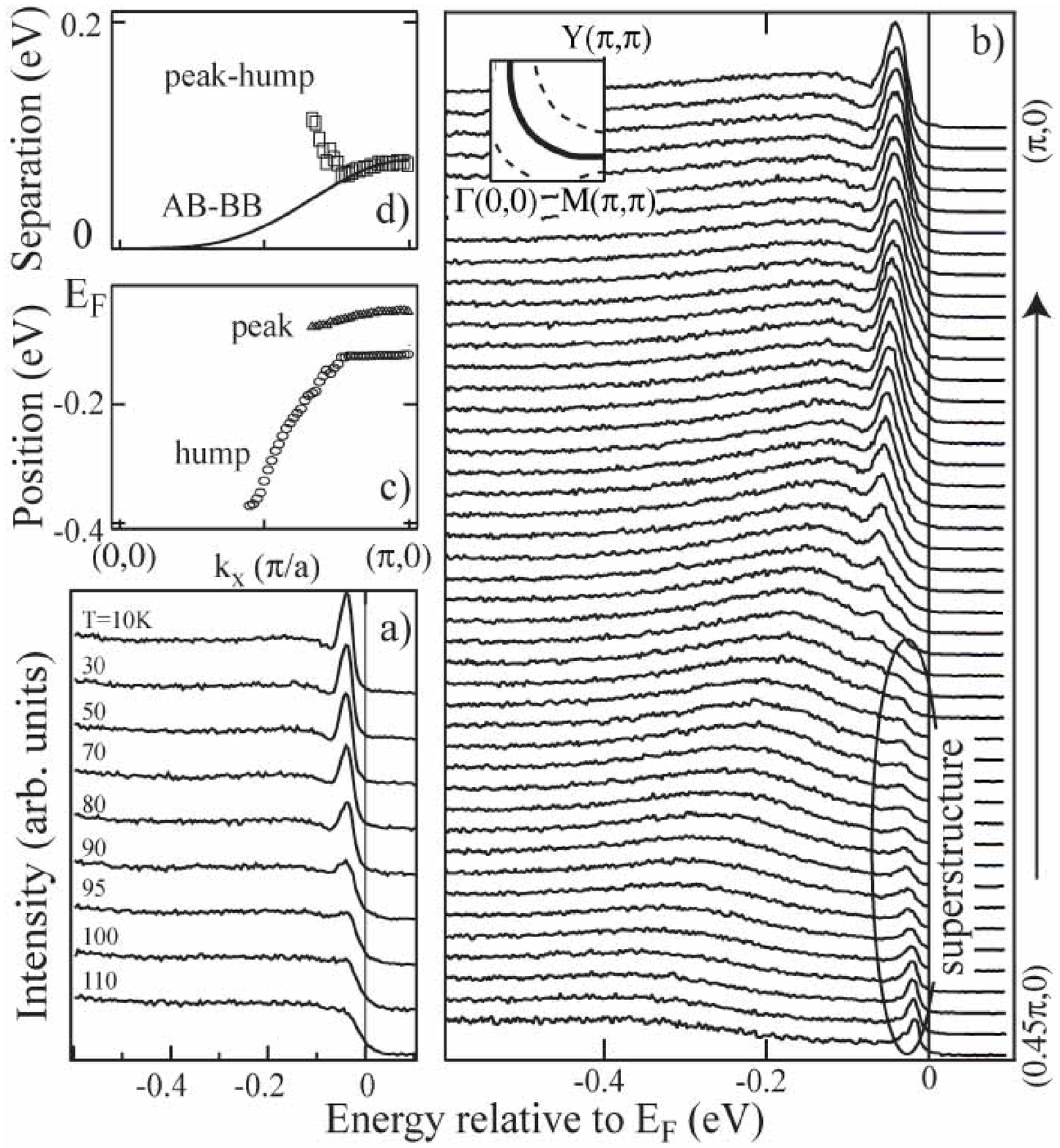}}
\caption[Dispersion of the PDH] {ARPES spectra for optimally doped
Bi2212 as a function of a) $T$ and b) momentum in the
superconducting state along $(0.45\pi,0)-(\pi,0)$. Features in the
circled region are due to the superstructures. c) The extracted
dispersions of the peak and hump based on the spectral maxima of
features. d) The calculated peak-hump spacing is compared with the
theoretical prediction for bilayer energy splitting in the
superconducting state.} \label{autonum1}
\end{figure}

\begin{figure}[t!]
\centerline{\includegraphics[width=3.3in]{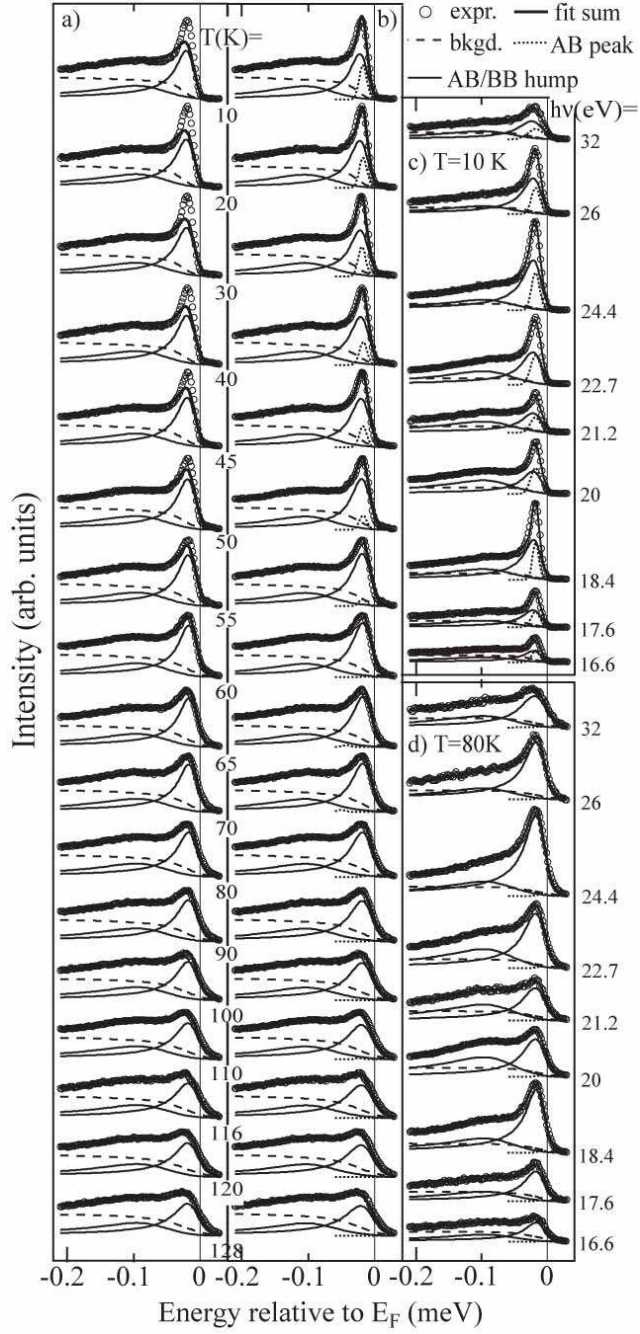}}
\caption[T-dependence Fit] {$T$-dependent ARPES spectra at
$(\pi,0)$ of OD65 in (a) fitted by 2-component model, (b) by
3-component model including $h\nu$-dependent spectra on Pb-OD65 at
(c) $T$=10 K and (d) $T$=80 K (see text). The AB hump lies lower
in energy than the BB hump. The width difference of the SCP is due
to resolution variations from 10 to 18 meV at different $h\nu$'s.}
\label{autonum2}
\end{figure}

\begin{figure}[t!]
\centerline{\includegraphics[width=3.5in]{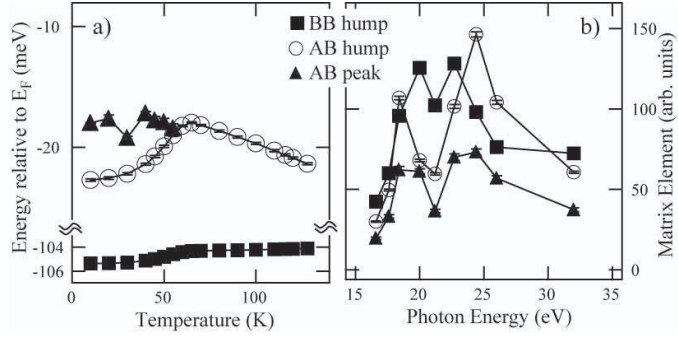}}
 \caption[hv-and T-dependence Fit] {The fitted a) $T$-dependent
 energy positions of the maxima of various features and (b) $h\nu$-dependent matrix elements of different
 spectral features. } \label{autonum3}
\end{figure}

\begin{figure}[t!]
\centerline{\includegraphics[width=3.5in]{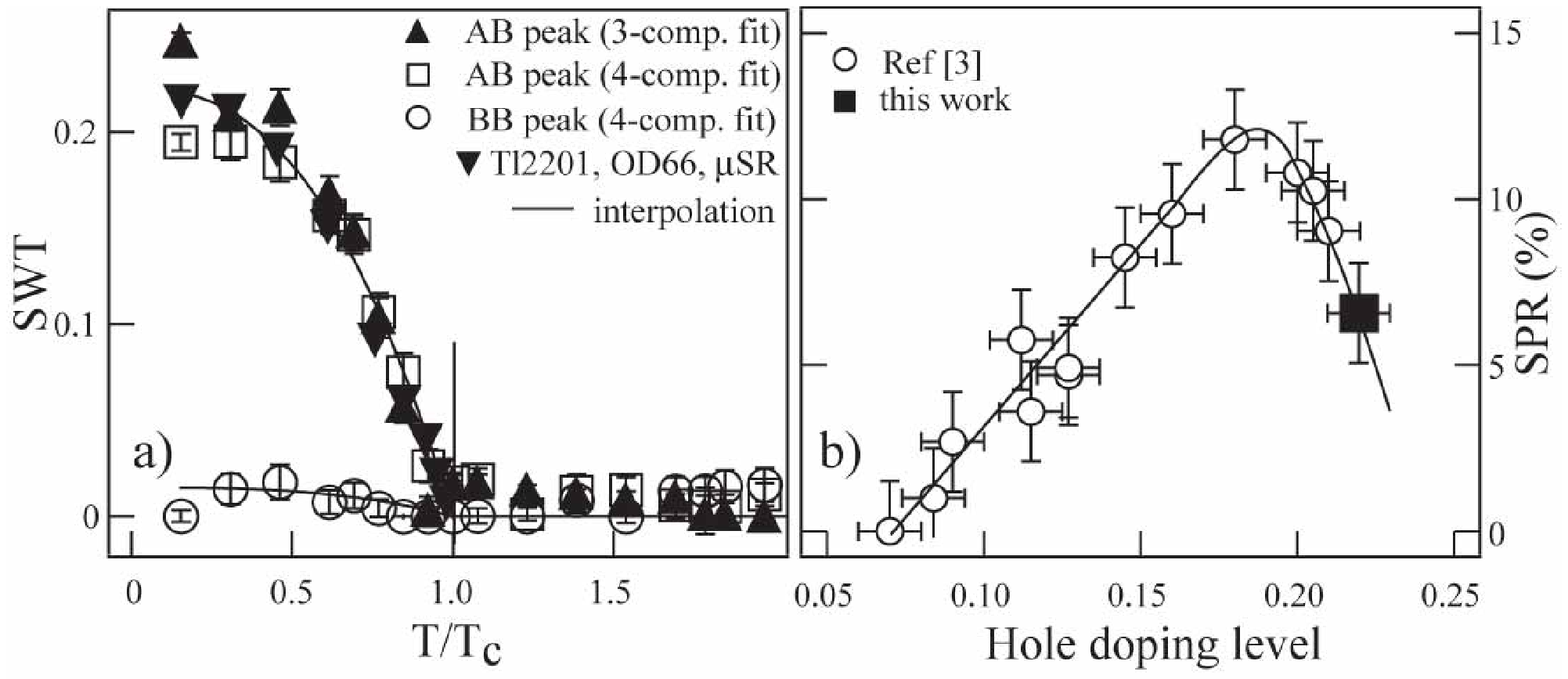}}%
\caption[SWT] {(a) The $T$-dependent SWT's given by 3-component
and 4-component fit are shown. The interpolations are based on a
trial function $SWT=SWT_{0}[1-(T/T_c)^{\zeta}]$\cite{TallonTl2201}
on the $T$-dependence of SWT (averaged) . The fitted value,
$\zeta\sim2.5$, agrees well with the $T$-dependent $\mu$SR result
on $Tl_2Ba_2CuO_{6+\delta}$ at the same doping level (OD66 Tl2201)
(after Ref.[18]). (b) The doping-dependence of SPR at $h\nu$=22.7
eV supplemented by the global fit result on OD65.}\label{autonum4}
\end{figure}

\begin{thebibliography}{}
\bibitem{BackgroundNote} We adopted a slightly different form of background approximation for a good match with the T-dependent EDC's. $B(\omega,T)=(1+b_1\omega+
 b_2\omega^2+b_3\omega^3)\cdot f(\omega-\sqrt{\Delta_b^{2}+\Delta_{sc}(T)^{2}},T+T_b)$, where $\Delta_{sc}(T)$ takes
 into account the superconducting gap (SCG) opening and the other parameters vary among experiments on different samples.
\bibitem{BBSamplitude} We define the
BBS amplitude in the experimental sense, i.e., the energy
separation between the maxima of AB and BB, which do not coincide
with the renormalized band energy position $\varepsilon$.
\end{thebibliography}
\end{document}